\documentclass[twocolumn, aps, showpacs] {revtex4}
\usepackage{graphicx}
\usepackage{amssymb}

\begin{document}
\draft

\title {Spin polarization and transition from metallic to insulating behavior in 2D systems}

\author {E. Tutuc, E.P. De Poortere, S.J. Papadakis and M. Shayegan}
\address{Department of Electrical Engineering, Princeton
University, Princeton, NJ 08544}

\begin{abstract}
We have made quantitative measurements of the spin polarization of two-dimensional (2D) GaAs (100) electrons and
GaAs (311)A holes, as a function of an in-plane magnetic field. The functional form of the in-plane
magnetoresistance is shown to be intimately related to the spin polarization. Moreover, for three different 2D
systems, namely GaAs (100) electrons, GaAs (311)A holes, and AlAs (411)B electrons, the temperature dependence of
the in-plane magnetoresistance reveals that their behavior turns from metallic to insulating {\it before} they are
fully spin polarized.
\end{abstract}

\maketitle

A 2D carrier system, long believed to have an insulating ground state \cite{gangoffour,bishop}, displays a
metallic behavior at finite temperature in certain ranges of densities \cite{MIpapers}. The response of the 2D
system to a parallel magnetic field ($B_{\|}$) may help understand this unexpected phenomenon
\cite{Mertes99,yoon00,Stergios:2D,tutuc,Okamoto99,Vitkalov,Murzin,Stergios:Sci,yaish,Stergios:symetry}: when
$B_{\|}$ is increased above a certain threshold value, $B_{I}$, the behavior of the 2D system changes to
insulating. Another effect of $B_{\|}$ is, of course, to create a Zeeman splitting between the spin-up and
spin-down states. This induces a partial spin polarization of the system so that two spin subbands with different
populations are formed. When $B_{\|}$ is increased above a characteristic field $B_{P}$, the 2D system becomes
fully spin polarized. The main goal of our work is to experimentally explore the relation between $B_{I}$ and
$B_{P}$.

\begin{figure*}
\centering
\includegraphics[scale=0.7]{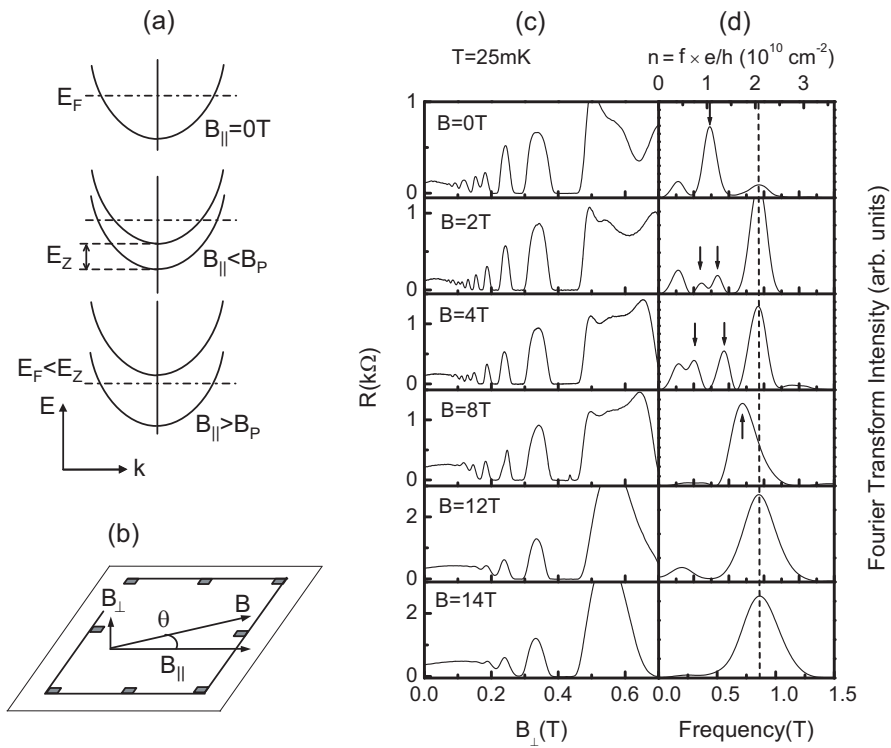}
\caption {\small{(a) Sketches of energy vs k-vector illustrating
the evolution of spin subbands with increasing parallel field. (b)
Experimental geometry. (c) Resistance vs $B_{\bot}$, for a GaAs
(100) 2D electron system with total density $n=2.05\times10^{10}$
cm$^{-2}$, at the indicated parallel fields. (d) Fourier
transforms of the traces in (c).}}
\end{figure*}

Here we report measurements of the fields $B_{P}$ and $B_{I}$ in
three different 2D systems. We first present, via Shubnikov - de
Haas (SdH) measurements in a nearly parallel field \cite{tutuc}, a
quantitative determination of the spin polarization of a dilute
GaAs 2D electron system as a function of $B_{\|}$. The data allow
us to find $B_{P}$. We then show that the sample magnetoresistance
(MR), measured as a function of a purely parallel field, reveals a
clear break at $B_{P}$, therefore providing an alternative method
for determining $B_{P}$. Our previous results, detailed elsewhere
\cite{tutuc}, showed a very similar trend for GaAs 2D holes. We
also present the temperature ($T$) dependence of the in-plane MR
to determine the field $B_{I}$ for three different
modulation-doped 2D systems: GaAs electrons, GaAs holes, and AlAs
electrons. In all three systems, $B_{I}$ is well below $B_{P}$
implying that the transition to insulating behavior always occurs
well before the full spin-polarization of the system.

\begin{figure}
\centering
\includegraphics[scale=0.65]{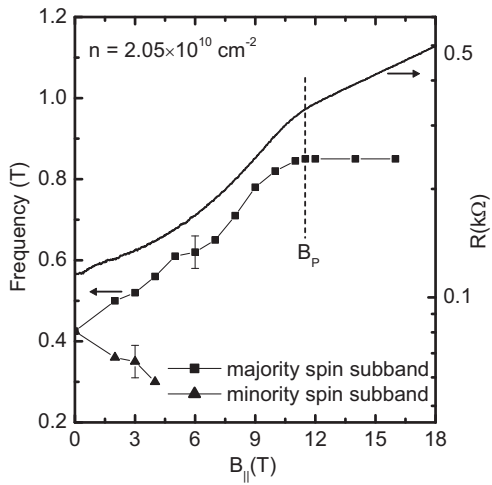}
\caption {\small{Summary of Fourier transform peak positions of
Fig. 1(d), as a function of applied magnetic field, along with the
in-plane magnetoresistance trace. The dashed vertical line marks
the field $B_{P}$ above which the 2D electrons becomes fully spin
polarized.}}
\end{figure}

Figures 1(a) and (b) illustrate how we use SdH measurements to determine the spin polarization. As shown
schematically in Fig. 1(a), application of a $B_{\|}$ to the 2D system separates the two spin subbands by the
Zeeman energy, $E_{Z}$. In our experiments [Fig. 1(b)] we apply a constant magnetic field ($B$), slowly rotate the
sample around $\theta=0^{\circ}$ to introduce a small $B_{\bot}$, and measure the sample resistance during the
rotation. We then Fourier analyze \cite{FT} the SdH oscillations induced by $B_{\bot}$, as their frequencies directly give
the two spin subbands' densities. We emphasize that in our measurements we are able to limit $B_{\bot}$ to
sufficiently small values so that $B_{\|} \cong B$ to better than 2\% during the sample rotation.

\begin{figure*}
\centering
\includegraphics[scale=0.65,angle=0]{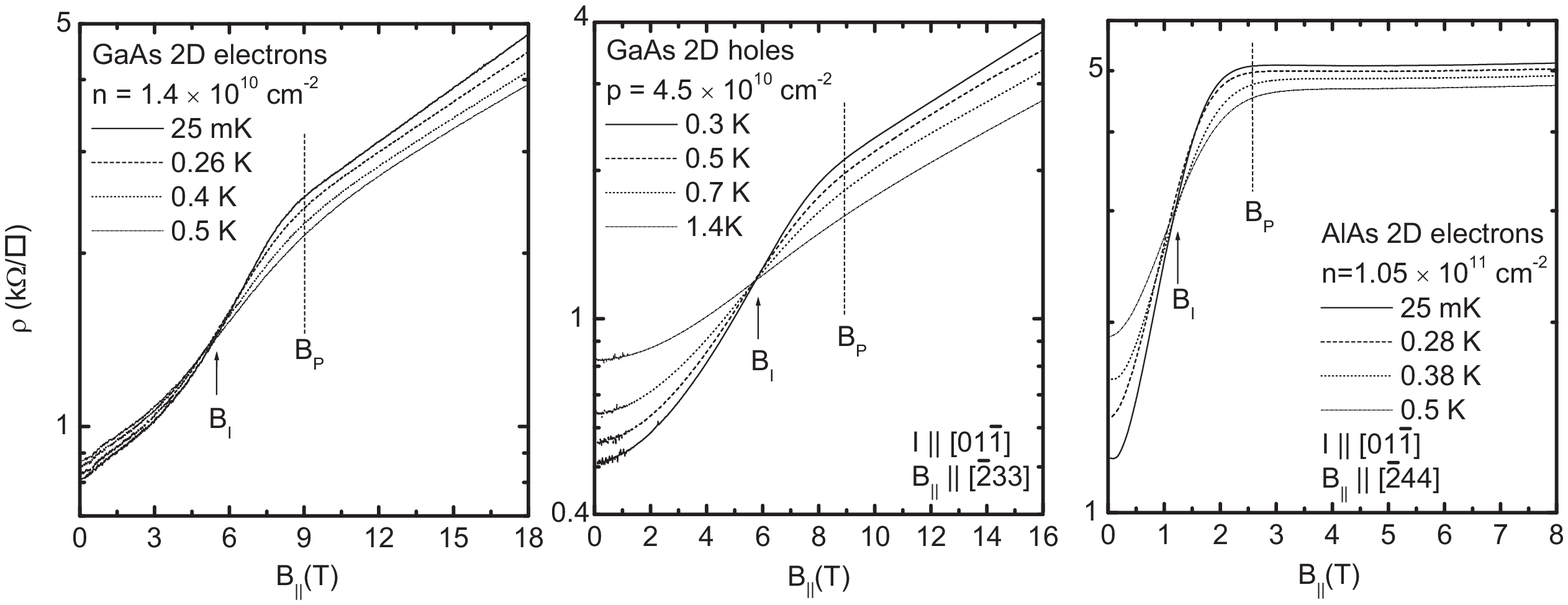}
\caption {\small{Temperature dependence of the in-plane MR for
three different 2D systems: (a) GaAs (100) electrons (b) GaAs
(311)A holes and (c) AlAs (411)B electrons. The density of each 2D
system is indicated in the respective panel. The fields $B_{I}$
and $B_{P}$ mark the onsets of insulating behavior and full spin
polarization, respectively.}}
\end{figure*}

In Fig. 1 (c) we show plots of resistance $R$ vs $B_{\bot}$, all
taken for a GaAs (100) 2D electron system at a fixed total
density, $n=2.05\times10^{10}$ cm$^{-2}$ \cite{R1}. The trace in
top box was taken with $\theta$ fixed at $90^{\circ}$ and
represents standard SdH measurements, i.e., $B_{\|}=0$T. The
Fourier transform of $R$ vs $1/B_{\bot}$, shown in the top box in
Fig. 1(d), exhibits two peaks, one at 0.85T and another at half
this value, at 0.42T. The 0.85T peak, multiplied by $(e/h)$, gives
the total density of the 2D electron system. The peak at 0.42T
stems from the spin-unresolved oscillations. The rest of the
traces shown in Fig. 1(c) were taken by rotating the sample at the
indicated constant $B$ applied almost parallel to the 2D plane.
With increasing $B_{\|}$, we observe a splitting of the lower
Fourier transform peak (0.42T) into two peaks. The frequencies of
these two peaks, multiplied by $(e/h)$, give the two spin subband
populations. We note that the two populations add up to the total
density of the sample. As $B_{\|}$ is increased, the majority spin
subband peak merges with the total density peak (0.85T) and the
minority spin subband peak moves to very low frequencies and is no
longer resolved \cite{R2}.

In Fig. 2 we summarize the positions of the Fourier transform
peaks corresponding to the majority and minority spin subbands as
a function of $B_{\|}$ for the case examined in Fig. 1. Above a
certain field $B_{P}$, the majority spin subband population
saturates at a value which corresponds to the total 2D density.
Therefore, $B_{P}$ marks the onset of full spin polarization. We
also measured the in-plane MR, by fixing $\theta$ at $0^{\circ}$
and recording the resistance as a function of a purely parallel
magnetic field. The MR trace, taken for $n=2.05\times10^{10}$
cm$^{-2}$, is also shown in Fig. 2. This trace exhibits a clear
break in the functional form of the MR as it changes from a strong
$\sim e^{B^{2}}$ dependence at low field to a weak $\sim e^{B}$
dependence at higher fields. The data in Fig. 2 demonstrate that
the break, i.e., the onset of simple exponential behavior of the
in-plane MR, coincides with the field $B_{P}$ above which the
spins are fully polarized. Remarkably, GaAs 2D holes exhibit a
similar functional form for the in-plane MR, and the field of full
spin polarization also coincides with the onset of the exponential
behavior \cite{yoon00,Stergios:2D,tutuc}.

Of particular relevance to the anomalous metallic behavior in 2D systems is the temperature dependence of the
in-plane MR. In Fig. 3 we show examples of such data for three different 2D systems: GaAs (100) electrons, GaAs
(311)A holes and AlAs (411)B electrons \cite{etienne}. The data, in agreement with previous results
\cite{Mertes99,yoon00,Stergios:2D,tutuc}, reveal that the metallic behavior observed at zero magnetic field turns
insulating above a certain field, $B_{I}$. The data also clearly show that $B_{I}$ is always smaller than $B_{P}$,
i.e., the transition to the insulating phase occurs \textit{before} the 2D system is fully spin polarized.

These observations provide strong experimental support for a link between the presence of two subbands with finite
populations and the metallic behavior. We suggest that a temperature dependent intersubband scattering mechanism
may be responsible for the metallic behavior.

To summarize, we have measured the spin subband populations of different 2D systems as a function of a parallel
magnetic field. We establish a direct connection between the functional form of the in-plane MR and the spin
polarization of the 2D system. Temperature dependence of in-plane MR data for three distinct 2D systems clearly
shows that each 2D system turns insulating \textit{before} it is fully spin polarized.

Our work was supported by ARO, DOE, NSF, and the von Humboldt Foundation. We thank R. Winkler for valuable
discussions.


\begin{thebibliography}{00}
\bibitem{FT}
In the range of magnetic field where the resistance displays SdH oscilations we first plot $R$ vs $1/B_{\bot}$. We
then substract a fitted linear background to center the oscillations around zero, multiply the signal with a
Hamming window, and then use a fast Fourier transfrom algorithm.
\bibitem{R1}
The total density was determined from the positions of integer and fractional quantum Hall states developed at
fields higher than 0.5T.
\bibitem{R2}
The peaks observed at very low frequencies ($<0.2$T) in Fig. 1(d) are Fourier transform artifacts.
\end{thebibliography}
\end{document}